\documentclass[aps,pra,superscriptaddress,amsmath,amssymb,preprintnumbers,showpacs,twocolumn]{revtex4}
\usepackage{amssymb}
\usepackage{graphicx}
\usepackage{bm}
\usepackage{color}

\newcommand{\ket}[1]{\left\vert #1\right\rangle}
\newcommand{\bra}[1]{\left\langle #1\right\vert}
\newcommand{\projector}[1]{\ket{#1}\bra{#1}}
\newcommand{\Trace}{\textrm{Tr}}

\newcommand{\fref}[1]{Fig. \ref{#1}}
\newcommand{\tr}{\textrm{Tr}}

\begin{document}

\title{Tripartite thermal correlations in an inhomogeneous spin-star system}

\author{F. Anz\`a}
\address{Dipartimento di Scienze Fisiche ed Astronomiche
dell'Universit\`{a} di Palermo, Via Archirafi 36, 90123 Palermo,
Italy}

\author{B. Militello}
\email{bdmilite@fisica.unipa.it}
\address{Dipartimento di Scienze Fisiche ed Astronomiche
dell'Universit\`{a} di Palermo, Via Archirafi 36, 90123 Palermo,
Italy}

\author{A. Messina}
\address{Dipartimento di Scienze Fisiche ed Astronomiche
dell'Universit\`{a} di Palermo, Via Archirafi 36, 90123 Palermo,
Italy}

\begin{abstract}
We exploit the tripartite negativity to study the thermal
correlations in a tripartite system, that is the three outer spins
interacting with the central one  in a spin-star system. We
analyze the dependence of such correlations on the homogeneity of
the interactions, starting from the case where central-outer spin
interactions are identical and then focusing on the case where the
three coupling constants are different. We single out some
important differences between the negativity and the concurrence.
\end{abstract}

\pacs{03.65.Ud, 03.67.Mn, 75.10.Jm}

%Entanglement and quantum nonlocality;
%Entanglement measures, witnesses, and other characterizations;
%Quantized spin models, including quantum spin frustration;

\maketitle

\section{Introduction}

The notion of thermal entanglement relies on the amount of
entanglement possessed by a physical system when it had undergone
a thermalization process that has brought it in a thermal state
\cite{ref:Arnesen2001}. In fact, in the presence of interaction
between different parts of a compound system, even after a
complete thermalization, the system can exhibit appreciable
quantum correlations since the Hamiltonian eigenstates are, in
general, entangled states. The establishment of a relation between
temperature and entanglement has quickly brought to the idea of
using the entanglement as an order parameter in quantum phase
transitions \cite{ref:Osterloh2002, ref:Osborne2002}. Moreover, it
has given a stronger stimulus of searching quantum correlations in
the macroscopic world
\cite{ref:Vedral2004,ref:Vedral2005,ref:Vedral2008}.

Thermal entanglement has been studied in connection with many
possible applications and in different physical systems: detailed
analysis in spin chains described by Heisenberg models has been
given \cite{ref:Gong2009} as well as in atom-cavity systems
\cite{ref:Wang2009} and in simple molecular models
\cite{ref:Pal2010}. The quantumness of correlations has been
singled out in thermalized spin chains \cite{ref:Werlang2010} also
in connection with non local effects \cite{ref:Souza2009}, and
application of thermal entanglement in optimal quantum
teleportation protocols has been proposed \cite{ref:Zhou2009}.

Spin systems have been studied in depth not only in the
neighbor-interaction configuration, leading to the Heisenberg
model, but also in star networks. This kind of systems can be
realized in many physical contexts, from Josephson Junctions
\cite{ref:Makhlin1999} to trapped ions \cite{ref:Cirac2000} to
solid state physics \cite{ref:Keane1998}. In 2004, Hutton and Bose
have analyzed a physical system made of a {\it central} spin
interacting with a set of $N$ {\it outer} spins at zero
temperature, bringing to light interesting properties  which are
immediately traceable back to the features of the ground state of
this spin-star network \cite{ref:Hutton2004}. They have shown that
the evenness or oddness of the number of outer spins determines
the law the entanglement amount scales with. The same spin
configuration, in a simplified version involving three outer spins
only, has been recently considered by Wan-Li {\it et al}
\cite{ref:Wan-Li2009} in the special case where all the
interactions between the central spin and the outer ones are
identical. Exploiting the high degree of symmetry of the system,
they concentrate on appropriate concurrences \cite{ref:Wootters}
to extract information about the presence of quantum correlations.
However, the disclosure of quantum correlations in tripartite
systems could require tools and concepts more adequate than the
simple concurrences. Sabin and Garcia-Alcaine have contributed to
solve this still very open problem\cite{ref:Fazio2008} introducing
the notion of tripartite negativity\cite{ref:Sabin2008}, which is
an effective tool to reveal the existence of quantum correlations
traceable back to the impossibility of separating any of the three
subsystems from the other two.

In this paper we consider a spin-star system where three outer
spins are coupled to the central one with different strengths, and
analyze thermal entanglement in different situations. In the next
section we present the model under scrutiny, which is
characterized by anisotropic spin-spin interactions due to the
absence of longitudinal $z$-$z$ couplings. In the third section we
consider the system in the homogeneous case, showing the presence
of sharp changes in the amount of entanglement at zero temperature
due to abrupt variations of the ground state. In the fourth
section we analyze the inhomogeneous model, focusing on two types
of inhomogeneity. Finally, in the last section some comments and
conclusive remarks are given.

\section{Physical scenario}

\begin{figure}
\begin{center}
\includegraphics[scale=0.7]{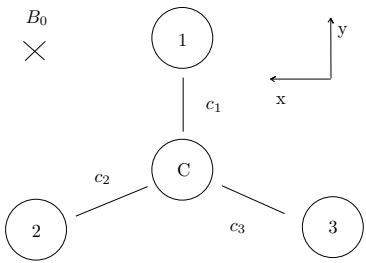}
\caption{The physical system. Spins labeled 1,2 and 3 are coupled
to the central (C) one. The magnetic field $B_0$ is ortogonal to
the plan where the four spins lie.} \label{fig:sys}
\end{center}
\end{figure}

\paragraph*{System and Hamiltonian -}
In this section we present the spin star-system we have analyzed,
which is pictured in \fref{fig:sys}. Numbers $1,2,3$ indicate the
three outer spin $1/2$, while the central one is indicated with
latin capital letter $C$. With $c_1, c_2$ and $c_3$ we indicate
the coupling constants of the central spin with the outer spin
$1$, $2$ and $3$, respectively. Moreover, the whole system is
immersed in a constant uniform magnetic field of modulus $B_0$ and
directed along the $z$-axis. The Hamiltonian model is then given
by (with $\hbar=1$):
\begin{equation} \label{eq:ham}
\mathbf{H} = \mathbf{H_1} + \mathbf{H_2} \; ,
\end{equation}
\begin{equation} \label{eq:ham1}
\mathbf{H_1} = -\frac{\gamma_s}{2} B_0 \sum_{i} \sigma_z^i =
\frac{\omega_0}{2} \sum_{i} \sigma_z^i \; ,
\end{equation}
\begin{equation} \label{eq:ham2}
\mathbf{H_2} = \sum_{i=1,2,3} c_i \left( \sigma_+^C \sigma_-^i +
\sigma_-^C \sigma_+^i \right) \; ,
\end{equation}
where $\sigma_z^i=\ket{+}_i\bra{+}-\ket{-}_i\bra{-}$,
$\sigma_+^i=\ket{+}_i\bra{-}$, $\sigma_-^i=\ket{-}_i\bra{+}$,
$\omega_0$ is the unperturbed Bohr frequency, and $c_i$'s are
coupling constants.

The term $\mathbf{H_1}$, describes the interaction of the system
with the magnetic field, while the second one, $\mathbf{H_2}$,
arises from the dipole-like interactions between the central spin
and the outer ones. The absence of $z$-$z$ interaction reflects a
certain degree of anisotropy.

Assuming that the system is in a thermal state, its density
operator takes the form:
\begin{equation} \label{eq:thermal}
\rho = \frac{e^{-\frac{\mathbf{H}}{k_BT}}}{\Trace \left(
{e^{-\frac{\mathbf{H}}{k_BT}}} \right)} \; =
\frac{e^{-\frac{\mathbf{H}}{k_BT}}}{Z} \; ,
\end{equation}
so that $\rho$ and $\mathbf{H}$ share the same eigenstates. The
Hamiltonian model given in Eq. (\ref{eq:ham}) takes into account
possible inhomogeneities, nevertheless it is surely of interest
studying it in the homogeneous case ($c_1=c_2=c_3=c$) which turns
out to be simpler from a mathematical point of view, paving the
way to the study of the more complicated inhomogeneous model.
Wan-Li {\it et al.} \cite{ref:Wan-Li2009} have analyzed the
homogeneous model taking advantage of the concurrence to quantify
entanglement. In addition, they have assumed a more general
anisotropic interaction. Since in real situations the construction
of a system with perfectly homogeneous interactions could be quite
difficult, in this paper we investigate the features of thermal
quantum correlations in the presence of inhomogeneity.

\paragraph*{Tripartite negativity -} \, Remarking that the analysis is performed on the outer
subsystem, which is made of three parts, we have the need to study
quantum correlations in tripartite systems. There are various
proposals of tripartite entanglement quantifiers
\cite{ref:TripQuantify-1,ref:TripQuantify-2,ref:TripQuantify-3,ref:TripQuantify-4}
and witnesses \cite{ref:TripWitnesses-1, ref:TripWitnesses-2,
ref:TripWitnesses-3}, the latter ones based on the key assertions
in \cite{ref:TripWitnesses}. Nevertheless, up to now and to the
best of our knowledge, there is not a definitive answer to the
request of quantifying tripartite entanglement
\cite{ref:Fazio2008}. For instance, it has been shown recently
that three-tangle \cite{ref:ThreeTangle} could be an improper tool
for such a purpose \cite{ref:DoesThreeTangle}. The situation is
much more complicated when we have to consider mixed states.
Indeed, all the recipes mentioned in
\cite{ref:TripQuantify-1,ref:TripQuantify-2,ref:TripQuantify-3,ref:TripQuantify-4,ref:ThreeTangle}
refer to pure states and fail when applied to non-pure states.
Since we are studying thermal correlations, therefore dealing with
highly mixed states, we need a tool to investigate quantum
correlations in mixed states of tripartite systems. Again, to the
best of our knowledge, the {\it tripartite negativity} introduced
by Sabin and Garcia-Alcaine \cite{ref:Sabin2008} is a good
quantity that allows to study quantum correlations in tripartite
systems, even in the case of mixed states. Other quantities are
very much related to the specific structure of the analyzed
system, since observables that assume values in a certain range
when the state is entangled are considered. Sabin and
Garcia-Alcaine apply bipartite negativity to all the three
possible bipartitions that can be singled out isolating one
subsystem and considering the other two as a whole; then they
consider the geometric mean of these three quantities. According
to the definition of negativity \cite{ref:Zyczkowski, ref:Vidal},
the partial negativity related to the bipartition $I - (JK)$ (in
which the two subsystems $J$ and $K$ are considered as a whole) is
given by
\begin{equation}
\mathcal{N}_{I-JK} = \sum_{i} |\sigma_i(\tau^{TI})| - 1 \quad ,
\end{equation}
where $\sigma_i(\tau^{TI})$ is the $i$-th eigenvalue of
$\tau^{TI}$, which is the partial transpose related to subsystem
$I$, of the total ($IJK$) density matrix. The parameter used to
study correlations in tripartite systems is then
\begin{equation} \label{eq:neggeom}
\mathcal{N}_{123} =
\sqrt[3]{\mathcal{N}_{1-23}\,\mathcal{N}_{2-13}\,\mathcal{N}_{3-12}}\;,
\end{equation}
which is the above mentioned {\it tripartite negativity}. In
\cite{ref:Sabin2008} the following properties have been proven:
\begin{itemize}
\item[i)] $\tau$ separable or simply bi-separable $\; \Rightarrow  \; \mathcal{N}_{123} = 0$;
\item[ii)] Invariance of $\mathcal{N}_{123}$ under LU operators;
\item[iii)] Monotonicity of $\mathcal{N}_{123}$ under LOCC operators;
\end{itemize}

Though $\mathcal{N}_{123}$ is not a sufficient condition to single
out tripartite entanglement, it is an effective tool to study
quantum correlations in tripartite systems. In fact, finding
$\mathcal{N}_{123} \neq 0$ implies that none of the three
subsystems is separable, hence revealing correlations involving
all the three subsystems. It is also important to note that
negativity $\mathcal{N}_{I-JK}$ does involve all the three spins,
and establishes the existence of a correlation between $I$ and the
entire couple made of $J$ and $K$, which is a very different
operation from tracing over one spin, say $K$, and evaluating the
degree of correlation between the other two, say $I$ and $J$.

\section{Homogeneous model} \label{sec:hom}

Let us consider our model in the special case $c_1=c_2=c_3=c$ that
we have already addressed as the homogeneous case. The complete
diagonalization of this Hamiltonian model is given in Appendix A.
In order to compute tripartite negativity we need first to find
explicit form of the outer-spin density operator $\rho_{123}$,
tracing over the degrees of freedom of the central spin.

\begin{figure}[h!]
\begin{center}
\includegraphics[scale=0.7]{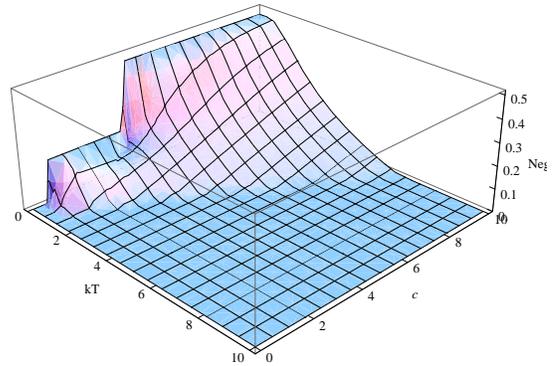}
\caption{Tripartite negativity $\mathcal{N}_{123}$ for the
homogeneous model as a functions of coupling constant ($c$) and
temperature ($k_BT$), both in units of $\omega_0$. Both the
diminishing for increasing temperature and the low-temperature
transitions with respect to $c$ are well visible.}
\label{fig:neghom}
\end{center}
\end{figure}

\begin{figure}[h!]
\begin{center}
\includegraphics[scale=0.5]{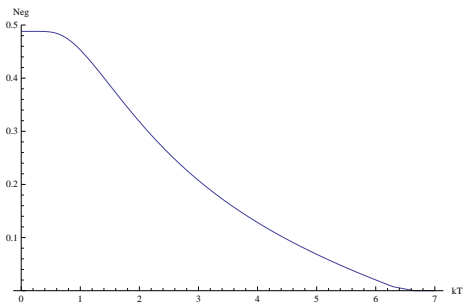}
\caption{Tripartite negativity $\mathcal{N}_{123}$ of the
homogeneous model versus $k_BT$, at $c=10 \omega_0$. Temperature
is in units of $\omega_0$. In the area where $k_BT<\omega_0 / 2$
we observe that the negativity is almost constant.}
\label{fig:Ccost}
\end{center}
\end{figure}

\begin{figure}[h!]
\begin{center}
\includegraphics[scale=0.7]{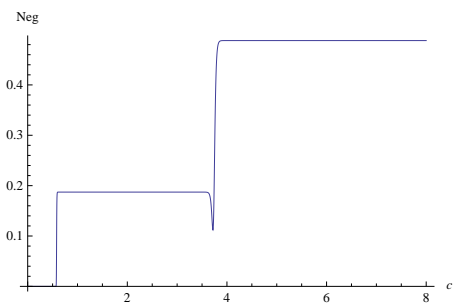}
\caption{Tripartite negativity $\mathcal{N}_{123}$ of the
homogeneous model versus coupling constants (in units of
$\omega_0$), at $k_BT=0.01 \omega_0$. Abrupt variations at
$c\approx 0.6\omega_0$ and $c\approx 3.7\omega_0$ are well
visible. Three plateaux, $A_1 \: (c < 0.6\omega_0), A_2 \:
(0.6\omega_0<c<3.72\omega_0), A_3 \: (c>3.72\omega_0)$ are quite
evident.} \label{fig:lowT}
\end{center}
\end{figure}

It is quite clear that because of homogeneity,
$\mathcal{N}_{1-23}=\mathcal{N}_{2-13}=\mathcal{N}_{3-12}$, so the
geometric mean reduces to one of these three quantities. In
\fref{fig:neghom} we plot $\mathcal{N}_{123}$ as a function of
temperature ($T$) and coupling constant ($c$). One can see some
interesting features of $\mathcal{N}_{123}$, for example its
diminishing as $T$ increases, and the presence of abrupt
transitions at low temperature due to fast variations of the
ground state. In \fref{fig:Ccost}, where $\mathcal{N}_{123}
(k_BT,c=6 \omega_0)$ is plotted, it is better shown this behavior
which is due to the fact that the more temperature increases the
more $\rho_{C123}$ approaches a maximally mixed states
($\frac{1}{16} \mathbb{I}$), making also $\rho_{123}$ maximally
mixed. A more interesting trend of $\mathcal{N}_{123}$ is observed
at low temperature where abrupt variations are present, as shown
in \fref{fig:lowT}. At $k_BT=0.01 \omega_0$ fast entanglement
variations are well visible at $c \approx 0.6 \omega_0$ and $c
\approx 3.7 \omega_0$. The reason for this occurrences is that in
each of these points the system undergoes a level crossing
involving its two lowest levels (ground and first excited states).
Let us call $A_1, A_2, A_3 \: $ the three {\it plateaux} of
\fref{fig:lowT}, corresponding to $(c < 0.6\omega_0) \:
(0.6\omega_0<c<3.72\omega_0)$ and $(c>3.72\omega_0)$,
respectively. The entanglement in these areas is the same of the
entanglement possessed by the ground state. In $A_1$ entanglement
is evidently zero and, in fact, the ground state of the whole
system is $\ket{g^{(A_1)}} = \ket{0000}$, corresponding to
$\rho_{123}^{(A_1)} = \ket{000} \bra{000}$ (with the notation
$\ket{\sigma_C\,\sigma_1\,\sigma_2\,\sigma_3}$). In regions $A_2$
and $A_3$ ground states of four spin system are $\ket{g^{(A_2)}}=$
$\left(\ket{0100} + \ket{0010} + \ket{0001} -\sqrt{3} \ket{1000}
\right)/\sqrt{6}$ and $\ket{g^{(A_3)}}=$ $\left( \left( \ket{0011}
+ \ket{0101} + \ket{0110} \right)\right.$ $-\left.\left(
\ket{1100} + \ket{1010} + \ket{1001} \right) \right) / \sqrt{6}$,
and the respective density matrix are $\rho_{123}^{(A_2)}=$
$\left( \ket{W} \bra{W} + \ket{000} \bra{000} \right)/2 $ and
$\rho_{123}^{(A_3)}=$ $\left( \ket{W} \bra{W} + |\tilde{W} \rangle
\langle \tilde{W} | \right)/2$, where $|W \rangle = \left(
\ket{100} + \ket{010} + \ket{001}\right)/\sqrt{3}$ and $|\tilde{W}
\rangle = \left( \ket{011} + \ket{101} +
\ket{110}\right)/\sqrt{3}$.

\section{Inhomogeneous model} \label{sec:unhomo}

Once we have analyzed the homogeneous model, we turn to the study
of the lack of homogeneity in the three interactions. For the sake
of simplicity we shall concentrate on two kinds of inhomogeneity
that we call type A ($c_1 = c$, $c_2 = c\,x$, $c_3 = c$) and type
B ($c_1 = c$, $c_2 = c\,x$, $c_3 = c\,x^2$), in both of which the
inhomogeneity is characterized by a suitable dimensionless real
parameter $x$. In both cases we study the thermal quantum
correlations of the tripartite outer subsystem as a function of
temperature ($T$), coupling parameter ($c$), and parameter of
inhomogeneity ($x$). It is worth noting that in the homogeneous
model the structure of eigenstates of $\mathbf{H}$ is independent
on the coupling constant ($c$), so that the features of the
correlations are essentially determined by the structure of the
eigenvalues only. Instead, in the inhomogeneous case the
eigenvalues are $c-$ and $x-$dependent and the eigenstates depend
on $x$, therefore the thermal correlations are affected both by
the crossing of levels and by the modification of the structure of
the eigenstates. We will see that while the change of the
eigenstates produces smooth changes of $\mathcal{N}_{123}$, on the
other hand a level crossing, especially at low temperature, can
produce a very sharp variation of the negativity.

\subsection{Model with inhomogeneity of type A}

In spite of the lack of some symmetry, the inhomogeneous model of
type A is still easily solvable and its diagonalization is given
in Appendix A.

\begin{figure}[h!]
\begin{center}
\includegraphics[scale=0.7]{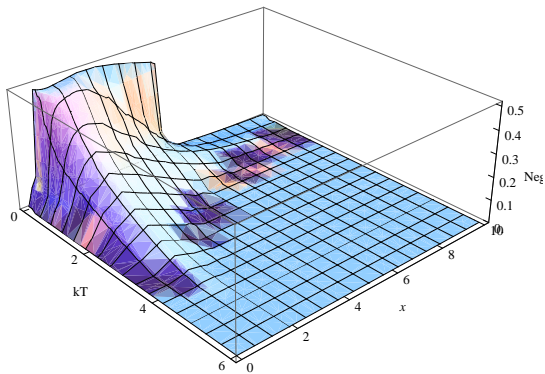}
\caption{Negativity $\mathcal{N}_{123}$ versus temperature
($k_BT$, in units of $\omega_0$) and inhomogeneity parameter
($x$), at $c=6\omega_0$, for the inhomogeneous model of type A.
The trend with respect to temperature is decreasing. The
dependence of $\mathcal{N}_{123}$ on the inhomogeneity parameter,
at low temperature, exhibits different trends: initially, smooth
changes, then, fast falling around $x\approx5.5$ and zero value
from there on.} \label{fig:UnHomACcost}
\end{center}
\end{figure}

Since one can see that in the Area $A_3$ of \fref{fig:lowT}
negativity assumes its greatest value, we start the analysis of
the inhomogeneous model of type A showing in
\fref{fig:UnHomACcost} the complete dependence of
$\mathcal{N}_{123}$ on temperature and inhomogeneity parameter for
$c=6 \omega_0$ (this specific value of $c$ belongs indeed to the
area $A_3$ of the homogeneous model). It is quite obvious that the
physical reason of the decreasing of $\mathcal{N}_{123}$ with
respect to $T$ is the same as for the homogeneous case, i.e. the
high degree of mixedness. At low temperature, in the $0<x<1$ area,
the smaller is $x$, the closer to zero is the negativity. In fact,
when $x$ is quite smaller than $1$, $c_2 = c \: x$ is much smaller
than $c_1 = c_3 = c$, and hence, roughly speaking, spin $2$ can be
considered almost decoupled, and then separable. Moreover, at $x_0
\approx 0.43$ this negligibility of the coupling between spin $C$
and spin $2$ is concomitant to a level crossing between the two
lowest energy levels, $E_2^{-} (c=6 \omega_0 , x) = - 3 \omega_0
\left( x + \sqrt{8+x^2} \right)$ and $E_4^{-} (c=6 \omega_0 , x) =
- (6 \sqrt{2+x^2} + 1)\,\omega_0$, that causes a sharp change at
$x=x_0$, where $E_2^{-} = E_4^{-}$.

Still, at low temperature, for $x$ quite greater than $1$,
$\mathcal{N}_{123}$ decreases as $x$ becomes increasingly grater
than $1$, because the coupling with spin $2$ becomes stronger than
the other two, making spin $1$ and $3$ almost separable from the
spin $2$. Around $x \approx 5.5$ a level crossing occurs. Because
of the diminishing of $\mathcal{N}_{123}$ for $x \gg 1$ and $x \ll
1$, there must be a maximum in the intermediate region.
Unexpectfully, this maximum is reached at $x=x_M \approx 2.46$,
instead of $x=1$. We think this is an important result because,
even though there is no big difference in the values of negativity
in a wide contour of the maximum (let's say from $x=0.5$ to
$x=5.5$). In fact, it is conceptually important that the maximum
of correlations is not reached in the homogeneous case, where the
high symmetry of the system could lead to the idea of stronger
correlations between its parts.

Evaluating the eigenvalues of Hamiltonian one can find that the
ground state in the nearby of this maximum is $\ket{\Psi_2^{-}}$,
which becomes $\ket{\Psi_M^{(A)}} \approx 0.2073 \left( \ket{0011}
+ \ket{1100}\right) + 0.2073 \left( \ket{1001} + \ket{0110}
\right) - 0.6435 \left( \ket{1010} + \ket{0101} \right)$ for
$x=x_{M}$. In this region, the values of $\mathcal{N}_{123}$ and
in particular the appearance of a maximum are essentially
determined by the dependence of the structure of the state
$\ket{\Psi_2^{-}}$ on the inhomogeneity parameter. The relevant
density operator is a mixture of two Werner-like states, and can
be written in the following way:
\begin{equation}
\rho^{2-}(x)=\frac{1}{2}\projector{w_1(x)}+\frac{1}{2}\projector{w_2(x)}\,,
\end{equation}
with
\begin{eqnarray}
\ket{w_1(x)}&=&
\aleph\left(\ket{011}+\ket{110}+\frac{\sqrt{8+x^2}-x}{2} \ket{101}\right)\,,\\
\ket{w_2(x)}&=&
\aleph\left(\ket{100}+\ket{001}+\frac{\sqrt{8+x^2}-x}{2}\ket{010}\right)\,,
\end{eqnarray}
where
\begin{equation}
\aleph = \left(2+\left((\sqrt{8+x^2}-x)/2\right)^2\right)^{-1/2}\,.
\end{equation}

\begin{figure}[t]
\begin{center}
\includegraphics[scale=0.7]{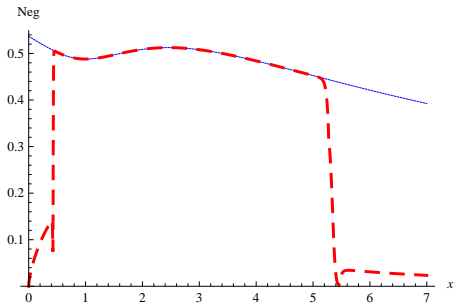}
\caption{Comparison of the Negativity ${\cal N}(\rho^{2-}(x))$
(solid blue line) with the Negativity of the thermal state (dashed
red line) evaluated for $c=6 \omega_0$ and $k_BT=0.01 \omega_0$,
both vs $x$. In the region $0.5<x<5.5$ (where $\ket{\Psi_2^{-}}$
is the ground state) the two curves coincide.}
\label{fig:NegativityPsiDueMeno}
\end{center}
\end{figure}

The relevant negativity can be evaluated and it turns out to be:
\begin{eqnarray}
\nonumber
&&{\cal N}(\rho^{2-}(x))=\frac{1}{2^{5/3}}\left\{-\frac{1}{(8+x^2)^{3/2}}\right.\\
\nonumber &\times&\left.\left[x+\sqrt{8+x^2}-\sqrt{2(20+x(x+\sqrt{8+x^2}))}\right]\right.\\
\nonumber &\times&\left.\left[x-3\sqrt{8+x^2}+8\sqrt{8+x^2}\right.\right.\\
&\times& \left.\left.
\left(|\lambda_1(x)|+|\lambda_2(x)|+|\lambda_3(x)|\right)
\right]^2\right\}^{1/3}
\end{eqnarray}
with $\lambda_i(x)$'s the roots of the algebraic equation in
$\lambda$:
\begin{eqnarray}
\nonumber
&&\left(-10x+x^3+(2+x^2)\sqrt{8+x^2}\right) \\
\nonumber
&&+ \left( -32 x - 4 x^3 - 48 \sqrt{8 + x^2} - 4 x^2 \sqrt{8 + x^2} \right) \lambda \\
\nonumber
&&+ \left(128 x + 16 x^3 - 384 \sqrt{8 + x^2} - 48 x^2 \sqrt{8 + x^2} \right) \lambda^2 \\
&&+ \left(1024 \sqrt{8 + x^2} + 128 x^2 \sqrt{8 + x^2} \right)\,\lambda^3\,=\,0\,,
\end{eqnarray}
where the coefficients, and hence the solutions, depend on $x$.
The behavior of ${\cal N}(\rho^{2-}(x))$ is shown in fig.
\ref{fig:NegativityPsiDueMeno}, where the negativity of the
thermal state for $c=6 \omega_0$ and $T\approx 0$ versus $x$ has
also been plotted. It is well visible that the behavior of ${\cal
N}(\rho^{2-}(x))$ perfectly reproduces the edge of the top visible
in fig. \ref{fig:UnHomACcost} in the low temperature region and
for $x$ belonging to the region where $\ket{\Psi_2^{-}}$ is the
ground state of the system.

\begin{figure}[t]
\begin{center}
\includegraphics[scale=0.4]{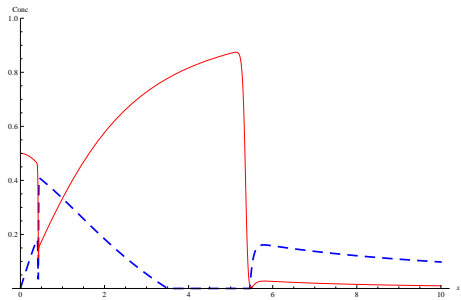}
\caption{The three concurrences $C_{12}=C_{23}$ (dashed blue line)
and $C_{13}$ (solid red line) vs $x$. Abrupt changes are very well
visible around $x=0.5$ and $x=5.5$, where the negativity exhibits
the same feature. For $x=1$ the three concurrences assume the same
value ($\approx 0.33$).} \label{fig:Concurrences}
\end{center}
\end{figure}

It could be interesting to compare the behavior of the tripartite
negativity with the values of the three concurrences related to
the three couples of spins that can be extracted tracing over one
of them: ${\cal C}_{23}={\cal C}(\tr_1\rho^{2-})$, ${\cal
C}_{12}={\cal C}(\tr_3\rho^{2-})$, ${\cal C}_{13}={\cal
C}(\tr_2\rho^{2-})$. Figure \ref{fig:Concurrences} shows that
around $x=0.5$ and $x=5.5$ there are abrupt changes in the values
of the concurrences (in the same points where the tripartite
negativity exhibits the same behavior), due to rapid changes of
the ground state. It is worth noting that in the region $3.5-5.5$
the concurrences $C_{12}$ and $C_{23}$ are vanishing while the
tripartite negativity (and hence all the three relevant bipartite
Negativity functions) is not. This apparent contradiction reflects
the very different meanings of these quantities. Indeed, for
example, while the concurrence $C_{12}$ measures the entanglement
between spins $1$ and $2$, without considering the spin $3$ that
has been preliminarily traced over, the negativity ${\cal
N}_{1-23}$ takes into account the correlations between spin $1$
and the whole couple made of the two spins $2$ and $3$. Therefore,
in that region, we can talk about correlations between spin $1$
and the couple $(2,3)$ without correlations between $1$ and any of
the other two spins, after the third has been traced over. We
think this is an important result, since it underlines the
difference between ${\cal N}$ and ${\cal C}$.

\begin{figure}[t]
\begin{center}
\includegraphics[scale=0.7]{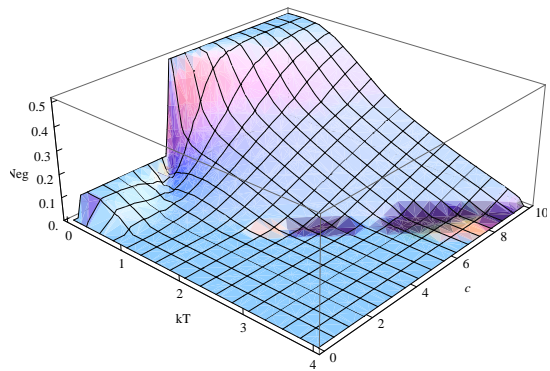}
\caption{Negativity $\mathcal{N}_{123}$ as a functions of
temperature ($k_BT$) and coupling parameter ($c$) at $x = 3$, for
the inhomogeneous model of type A, both $k_BT$ and $c$ are in
units of $\omega_0$. With respect to $c$, at low temperature,
there are abrupt changes traceable back to the occurrence of level
crossings.} \label{fig:UnHomAXcost}
\end{center}
\end{figure}

Figure \ref{fig:UnHomAXcost} shows the complete dependence of
$\mathcal{N}_{123}$ on temperature and coupling parameter, for
$x=3$. We observed (by performing other plots, which we are not
reporting here) that the qualitative behavior of
$\mathcal{N}_{123}$ is independent on the specific value of $x$,
provided it is neither too small nor too big. It is quite evident
that the trend with respect to temperature is the usual one. At
low temperature the dependence of $\mathcal{N}_{123}$ on $c$
exhibits only the step trend due to level crossings. The highest
step involves the energy levels $E_4^{-}(c,x=3) = - c \sqrt{11} -
\omega_0$ and $E_2^{-}(c,x=3) = - \left( 3 + \sqrt{17} \right) c/2
$ (see Appendix A), and it is quite easy to find that the crossing
occurs at $c_t \approx 4.08 \omega_0$ where $E_4^{-} = E_2^{-}$
while $E_4^{-} < E_2^{-}$ for $c<c_t$ and $E_2^{-} < E_4^{-}$ for
$c>c_t$.

\begin{figure}[t!]
\begin{center}
\includegraphics[scale=0.7]{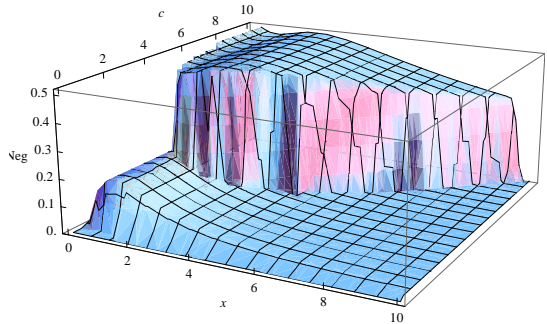}
\caption{Negativity $\mathcal{N}_{123}$, at low temperature
($k_BT=0.01 \omega_0$), versus coupling parameter ($c$, in units
of $\omega_0$) and inhomogeneity parameter ($x$), for the
inhomogeneous model of type A. Both smooth decreasing and fast
variations occur with respect to $x$. Dependence on $c$ exhibits a
stepwise behavior.} \label{fig:UnHomATcost}
\end{center}
\end{figure}

To conclude the analysis of the inhomogeneous model of type A, in
\fref{fig:UnHomATcost} we have plotted the complete dependence of
$\mathcal{N}_{123}$ on coupling and inhomogeneity parameters,
fixing the value of temperature at $k_BT =0.01 \omega_0$. The
dependence on the coupling parameter is qualitatively the same as
for the previous case (see \fref{fig:UnHomAXcost}). Indeed, we
observe a step trend caused by a level crossing between the two
lowest energy levels. As a function of $x$, $\mathcal{N}_{123}$
exhibits both smooth decreasing, due to increasing separability of
spins $1$ and $3$, and fast variations traceable back to a level
crossing amongst the two lowest energy levels $E_2^{-}$ and
$E_4^{-}$. In fact, we verified that the big step observed in
\fref{fig:UnHomATcost} corresponds to the locus of points of the
plane $(c,x)$ in which the two lowest levels have the same energy.
Thanks to the simple form of the eigenvalues of our model, one can
find a simple analytical expression of this curve:

\begin{equation}
c(x) = \frac{x^2 + 4 + \alpha (x)}{4x} + \frac{1}{2}
\sqrt{\frac{x^2+5+\alpha (x)}{2}} \quad ,
\end{equation}
where $\alpha (x)$ is
\begin{equation}
\alpha (x) = \sqrt{16 + 10x^2 + x^4} \quad .
\end{equation}

\subsection{Model with inhomogeneity of type B}

The second type of inhomogeneity we decided to study is
characterized by the coupling constants $c_1=c$, $c_2=c\:x$ and
$c_3= c\:x^2$. Also for this model we have considered the
dependence on $T$, $c$ and $x$. Surprisingly, what we have found
is that the behavior of negativity in the presence of
inhomogeneity of type B is not different from what we have found
in the presence of inhomogeneity of type A. The diminishing for
increasing temperature, the abrupt changes at low temperature in
connection with level crossings and the presence of a maximum for
$x\not=1$ are all present also in this model of inhomogeneity.
Since for type B we do not have an explicit analytical
diagonalization of the Hamiltonian, we also miss an analytical
expression of the state of the system in the nearby of the
maximum. In fact, we have only numerical results, according to
which, for example, for $T=0.01\,\omega_0/k_B$ and $c=6\,\omega_0$
the maximum of negativity occurs at $x=x_M\approx 1.91$, where the
ground state of the system is approximately: $\ket{\Psi_{M}^{(B)}}
= 0.292 \left( \ket{1100} - \ket{0011} \right) + 0.471 \left(
\ket{1001} - \ket{0110} \right) + 0.439 \left( \ket{1010} -
\ket{0101} \right)$.

\section{Conclusions}

In this paper we have analyzed the thermal correlations in a
spin-star system consisting of a central spin interacting with
three outer ones, all immersed in a magnetic field. In the absence
of a tripartite entanglement measure, we have decided to exploit
the tripartite negativity, which is at least able to put in
evidence the lack of separability and simple bi-separability of a
mixed quantum state of a tripartite system.

We started from considering the homogeneous model in which the
three peripheral spins interact in the same way with the central
one, showing the appearance of a significant degree of
inseparability revealed by a non vanishing tripartite negativity
${\cal N}_{123}$. Typical behavior of thermal correlations
reflects the behavior of ${\cal N}_{123}$, which decreases to
zero-value as temperature increases and, instead, exhibits both
appreciable values and abrupt changes at very low temperature.
These occurrences show in a very clear way the role of thermal
entanglement mediator played by the central spin. Further we have
considered the presence of inhomogeneity, meant as differences in
the coupling constants between the outer spins and the central
one, focusing on two special cases. In the first case
(inhomogeneity of type A) two spins interact in the same way with
the central one and the third has a different coupling constant.
In the second case (inhomogeneity of type B) all the three outer
spins are characterized by different coupling strengths. Though
some differences are visible, the qualitative behaviors of ${\cal
N}_{123}$ are quite similar for the two inhomogeneous models and
even for the homogeneous one. This suggests the idea that the
thermal quantum correlations mediated by the central spin are not
very much damaged by a certain lack of homogeneity that could
characterize a more realistic situation, provided the degree of
inhomogeneity is not high.

A remarkable point is that we have singled out the presence of
maxima of tripartite negativity corresponding to inhomogeneous
models, i.e. for the values of the inhomogeneous parameter quite
larger than $1$. Though the differences between a maximum value of
negativity and its values in the nearby (even up to the
homogeneous case, i.e. up to $x=1$) is not very big, we think that
it is conceptually important the fact that a higher degree of
symmetry in the system does not guarantee a higher degree of
correlations between all its parts.

Another important point is that we have found some regions of the
parameter space where it happens that two concurrences, say ${\cal
C}_{12}$ and ${\cal C}_{23}$, are vanishing, meaning that there is
no entanglement between $1$ and $2$ and between $2$ and $3$, while
the relevant negativity ${\cal N}_{2-13}$ is non-vanishing,
meaning that there is a correlation between $2$ and the couple
made of $1$ and $3$. This fact points out in a very transparent
way the different meanings of concurrence and negativity.

\appendix

\section{}
In this appendix we give eigenvalues and eigenvectors of the
Hamiltonian of the inhomogeneous model of type A ($c_1=c$, $c_2=c
\: x$, $c_3 = c$), as functions of $x>0$, $c>0$, $\omega_0$. The
special case $x=1$ gives the solutions for the homogeneous model.

\begin{widetext}
\begin{displaymath}
\begin{array}{ll}
\textrm{Energies} & \textrm{Eigenstates} \\

E_1^{\pm} = \pm c\: x\,, & \ket{\Psi_{1}^{\pm}} = \frac{1}{2} \left[ \left(\ket{0011} \pm \ket{1100}\right) - \left(\ket{0110} \pm \ket{1001}\right) \right]\,,  \\

E_2^{\pm} = \pm \frac{c}{2} \left[x + \left(8+x^2\right)^{\frac{1}{2}}\right]\,, & \ket{\Psi_{2}^{\pm}} = \frac{1}{K_1} \: \left[ \left(\ket{0011} \pm \ket{1100}\right) + \left(\ket{0110} \pm \ket{1001}\right)\right. \\
& \left.+ \frac{\sqrt{8+x^2} - x}{2} \left( \ket{0101} \pm \ket{1010} \right) \right]\,,  \\

E_3^{\pm} = \pm \frac{c}{2} \left[x-\left(8+x^2\right)^{\frac{1}{2}} \right]\,, & \ket{\Psi_{3}^{\pm}} = \frac{1}{K_1} \: \left[ \left(\ket{0011} \pm \ket{1100}\right) + \left(\ket{0110} \pm \ket{1001} \right)\right.\\
& \left.- \frac{\sqrt{8+x^2}+x}{2} \left( \ket{0101} \pm \ket{1010} \right) \right]\,, \\

E_4^{\pm} = \pm \left[c \left(2+x^2\right)^{\frac{1}{2}}+\omega_0 \right]\,, & \ket{\Psi_{4}^{\pm}} = \frac{1}{K_2} \left[\sqrt{2+x^2} \ket{0111}\right. \\
& \left.\pm \left( \ket{1011} + x\ket{1101} + \ket{1110} \right) \right]\,,\\

E_5^{\pm} = \pm \left[c \left(2+x^2\right)^{\frac{1}{2}} - \omega_0 \right]\,, & \ket{\Psi_{5}^{\pm}} = \frac{1}{K_2} \left[\left( \ket{0100} + x \ket{0010} + \ket{0001} \right)\right.\\ &\left. \pm \sqrt{\left( 2+x^2 \right)} \ket{1000} \right]\,,\\

E_6 = -\omega_0\,, & \ket{\Psi_{6}^A} = \frac{1}{K_3} \left[ \frac{1}{x} \ket{0001} + \ket{0010} - \left( \frac{1}{x} + x \right) \ket{0100} \right] \,,  \\ & \ket{\Psi_{6}^{B}} = \frac{1}{\sqrt{1+x^2}} \left( \ket{0010} - x \ket{0001} \right)\,,   \\

E_7 = \omega_0\,, & \ket{\Psi_{7}^A} = \frac{1}{K_3} \left[ \frac{1}{x} \ket{1011} + \ket{1101} - \left( \frac{1}{x} + x \right) \ket{1110} \right] \,,  \\ & \ket{\Psi_{7}^{B}} = \frac{1}{\sqrt{1+x^2}} \left( \ket{1101} - x \ket{1011} \right)\,,  \\

E_8 = -2\omega_0\,, & \ket{\Psi_{8}} = \ket{0000}\,, \\
E_9 = 2\omega_0\,, & \ket{\Psi_9} = \ket{1111}\,, \\
\end{array}
\end{displaymath}
\end{widetext}
where $K_1$, $K_2$ and $K_3$ are:
\begin{displaymath}
{K_1}^2 = 4 + 2 \left( \frac{\sqrt{8+x^2} - x}{2} \right)^2\,,
\end{displaymath}
\begin{displaymath}
{K_2}^2 = 2\left(2+x^2\right)\,,
\end{displaymath}
\begin{displaymath}
{K_3}^2 = \frac{2}{x^2} + 3 + x^2\,.
\end{displaymath}

\end{document}